\documentclass[a4paper,12pt]{article}
\usepackage{hyperref}
\usepackage{slashed}
\usepackage{amsfonts}
\usepackage{multirow}
\usepackage{mathrsfs}
\usepackage{graphicx}
\usepackage{amsmath}
\usepackage{amssymb}
\usepackage[numbers, square, comma, sort&compress]{natbib}
\usepackage{bm}
\usepackage{bbm}
\usepackage{color}
\usepackage{geometry}
\usepackage[utf8]{inputenc}
\newcommand{\beq}{\begin{equation}}
\newcommand{\eeq}{\end{equation}}
\newcommand{\bqa}{\begin{eqnarray}}
\newcommand{\eqa}{\end{eqnarray}}
\newcommand{\GG}{\ensuremath \text{F}}
\newcommand{\M}{\ensuremath \text{M}}

\begin{document}

\begin{titlepage}
\begin{flushright}
TUM-HEP-1183/19\\ [0.2cm]
March 11, 2019
\\  
\end{flushright}

\vskip 25mm

\begin{center}

\Large\bf{\large\bf Master integrals of a planar double-box family for top-quark pair production}
\end{center}

\vskip 8mm

\begin{center}

{\bf Long-Bin Chen$^{1}$, Jian Wang$^2$}\\
\vspace{10mm}
\textit{$^1$School of Physics and Electronic Engineering, Guangzhou University, Guangzhou 510006, China}\\
\textit{$^2$Physik Department T31,  Technische Universit\"at M\"unchen, James-Franck-Stra\ss e~1,
D--85748 Garching, Germany} \\ 
\vspace{5mm}
\textit{E-mail:} \texttt{{\href{chenlb@gzhu.edu.cn}{chenlb@gzhu.edu.cn}, \href{j.wang@tum.de}{j.wang@tum.de}}}

\end{center}

\vspace{10mm}

\begin{abstract}
We calculate analytically the master integrals of a planar double-box family for top-quark pair production using the method of differential equations. With a proper choice of the bases, the differential equations can be transformed to the $d$-log form.
The square roots of the kinematic variables in
the differential equations can be rationalized by 
defining two dimensionless variables. 
We find that all the boundary conditions
can be fully fixed either by  simple integrals
or regularity conditions at some special kinematic points.
The analytic results for thirty-three master integrals at general kinematics are all expressed in terms of multiple polylogarithms up to  transcendental  weight four.

\end{abstract}

\end{titlepage}

\section{Introduction}
The top-quark pair production is one of the most important processes 
at a hadron collider, such as the LHC. It has a very large production rate and
the decay of the top quarks gives rise to several jets or leptons, which can be considered as an important background  
in the search of new physics. Moreover, this process can also be used to determine the top-quark mass, 
the strong coupling constant $\alpha_s$ and the gluon parton  distribution functions.
As such, it is important to have a precise understanding of this process. 
So far, the LHC experiment has accumulated a large number of data at 13 TeV.
The cross section of the top-quark pair production has been measured with a precision
comparable to the most precise theoretical predictions \cite{Aaboud:2016pbd,Sirunyan:2017uhy}.

The total cross sections and differential distributions of the top-quark pair production have been calculated 
up to next-to-next-to-leading order (NNLO) \cite{Baernreuther:2012ws,Czakon:2013goa,Czakon:2015owf,Catani:2019iny,Behring:2019iiv}. 
As a part of the calculation, the two-loop virtual corrections have been evaluated numerically \cite{Baernreuther:2013caa,Chen:2017jvi}. 
However, the analytic results of the two-loop diagrams
are still valuable in order to provide a fast and stable evaluation 
of the virtual corrections and to understand the structure
of massive loop integrals, which are usually much more 
complicated than the massless ones.
Some analytic results have already been obtained in refs. \cite{Bonciani:2008az,Bonciani:2009nb,Bonciani:2010mn,Bonciani:2013ywa}.

In the calculation of the two-loop Feynman diagrams, all the integrals
can be reduced to a set of master integrals, e.g. using integration-by-parts (IBP) identities.
The master integrals for the top-quark pair production have been widely studied. 
In ref.\cite{vonManteuffel:2013uoa}, the authors calculated the master integrals for the light fermionic two-loop QCD corrections to top-quark pair production in the gluon fusion channel. 
In refs.\cite{Mastrolia:2017pfy,DiVita:2018nnh}, the master integrals for the NNLO QED corrections to $\mu e$ scattering have been obtained.
Some of these master integrals are also applicable to top-quark pair production. 
Recently, the planar double-box integrals for top-quark pair production with a closed top-quark loop have been 
calculated \cite{Adams:2018bsn,Adams:2018kez}, of which the results contain elliptic integrals. 
The method of differential equations \cite{Kotikov:1990kg,Kotikov:1991pm} has played an important role in the above computations.

\begin{figure}[h]
\begin{center}
\includegraphics[scale=0.5]{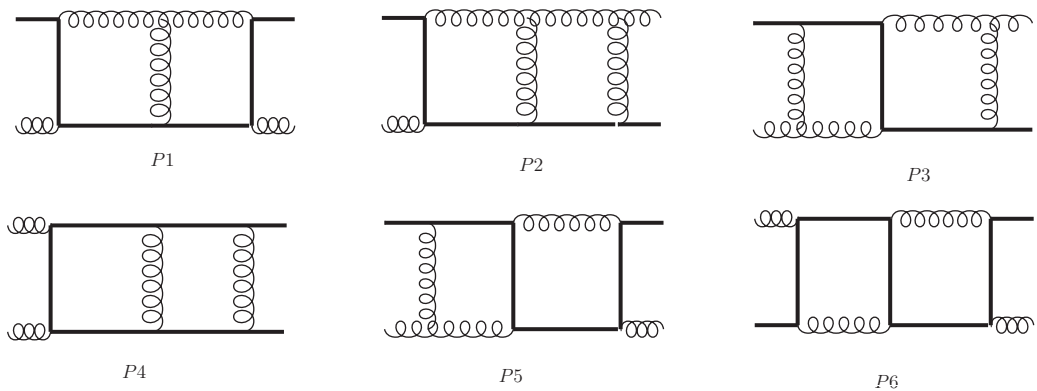}
\caption{Analytically unknown planar double-box integrals for top-quark pair production.}
\label{unknown1}
\end{center}
\end{figure}

\begin{figure}[h]
\begin{center}
\includegraphics[scale=0.5]{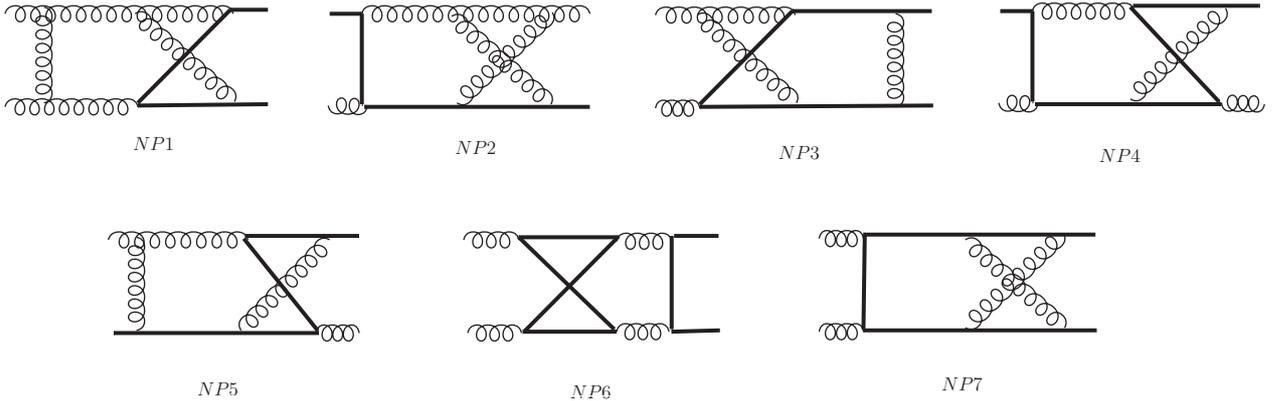}
\caption{Analytically unknown non-planar double-box integrals for top-quark pair production.}
\label{unknown2}
\end{center}
\end{figure}

We have examined all the Feynman integrals relevant to the NNLO corrections to top-quark pair  production, and found  six planar  and seven non-planar double-box integral families  remaining to be calculated analytically. 
We shown the corresponding planar  and non-planar double-box integrals in Fig.\ref{unknown1} and Fig.\ref{unknown2}, respectively.  

\begin{figure}[t]
\begin{center}
\includegraphics[scale=0.35]{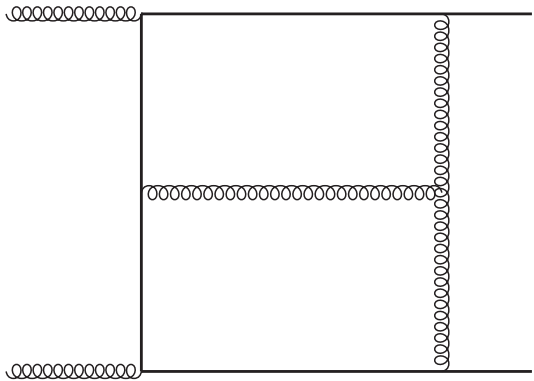}
\caption{A planar double-box Feynman diagram for top-quark pair production in gluon-gluon fusion (corresponding to $g(k_1) g(k_2)\to t(k_3)\bar{t}(k_4)$ or 
$g(k_1) g(k_2)\to t(k_4)\bar{t}(k_3)$).}
\label{midiag1}
\end{center}
\end{figure}

In this paper, we calculate one of the analytically unknown planar double-box integrals, i.e., $P1$ in Fig.\ref{unknown1},  for top-quark pair production.
We show its topology individually in Fig.\ref{midiag1}.
It turns out that the differential equations for all the master integrals in this family can be transformed to the $d$-log form
after choosing a proper basis.
In addition, the square roots in the logarithms 
can be rationalized by defining new dimensionless variables.
As a consequence, all the master integrals in this family can be written in terms of multiple polylogarithms. 
Note that the integral family we calculate does not contribute to the leading color results in \cite{Bonciani:2010mn}.
The other analytically unknown planar double-box integrals involve elliptic functions and will be discussed elsewhere.

The rest of this paper is organized as follows. In section \ref{sec2} we present the canonical basis of the integral family and their corresponding differential equations in the $d$-log form. We discuss the determination of boundary conditions in section \ref{sec3}. Conclusions are given in section \ref{sec4}. The analytic results as well as the rational matrices are provided in ancillary files along with this paper.

\section{Canonical basis and differential equations}
\label{sec2}

As shown in Fig.\ref{midiag1}, the planar double-box Feynman integrals in the family we consider can be formulated as
\bqa
I_{n_1,n_2,\ldots,n_{9}}=\int{\mathcal D}^d q_1~{\mathcal D}^d q_2\frac{D_8^{-n_8}~D_9^{-n_9}}{D_1^{n_1}~D_2^{n_2}~D_3^{n_3}~D_4^{n_4}~D_5^{n_5}~D_6^{n_6}~D_7^{n_7}}
\label{def}
\eqa
with the propagators given by
\bqa
D_1&=&q_1^2,D_2=q_2^2,D_3=(q_1+q_2)^2,D_4=(q_1-k_3)^2-m^2,D_5=(q_1+k_1-k_3)^2-m^2,\nonumber\\
D_6&=&(q_2-k_1-k_2+k_3)^2-m^2,D_7=(q_2-k_1+k_3)^2-m^2,\nonumber\\
D_8&=&(q_1-k_1-k_2+k_3)^2-m^2,D_9=(q_2-k_3)^2-m^2.\nonumber
\label{int}
\eqa
The measure of the integral is defined as
\beq
{\mathcal D}^d q_i = \frac{m^{2\epsilon}}{\pi^{D/2}\Gamma(1+\epsilon)}  d^d q_i~, \ \quad
d=4-2\epsilon.
\eeq
The external gluons and top-quarks are on-shell, i.e., $k_1^2=0,k_2^2=0,k_3^2=m^2$ and $k_4^2=(k_1+k_2-k_3)^2=m^2$.
The Mandelstam variables can be written as
\beq
s=(k_1+k_2)^2\,, \qquad t=(k_1-k_3)^2\,, \qquad u=(k_2-k_3)^2 
\eeq
with $s+t+u=2m^2$.
Notice that in the definition of the integral family in eq.(\ref{def}), the denominator $D_i,i=1,\cdots, 7$ are specified by
the propagators in Fig.\ref{midiag1}.
The denominators $D_8$ and $D_9$ are chosen 
in such a way that the integrals are symmetric
under the replacement $(k_1\leftrightarrow k_2, k_3\leftrightarrow k_4, q_1\leftrightarrow q_2)$.
And more importantly, they are vanishing 
in the infrared limit $q_i\to 0$ so that the results have better infrared behavior.

\begin{figure}[t]
\begin{center}
\includegraphics[scale=0.3]{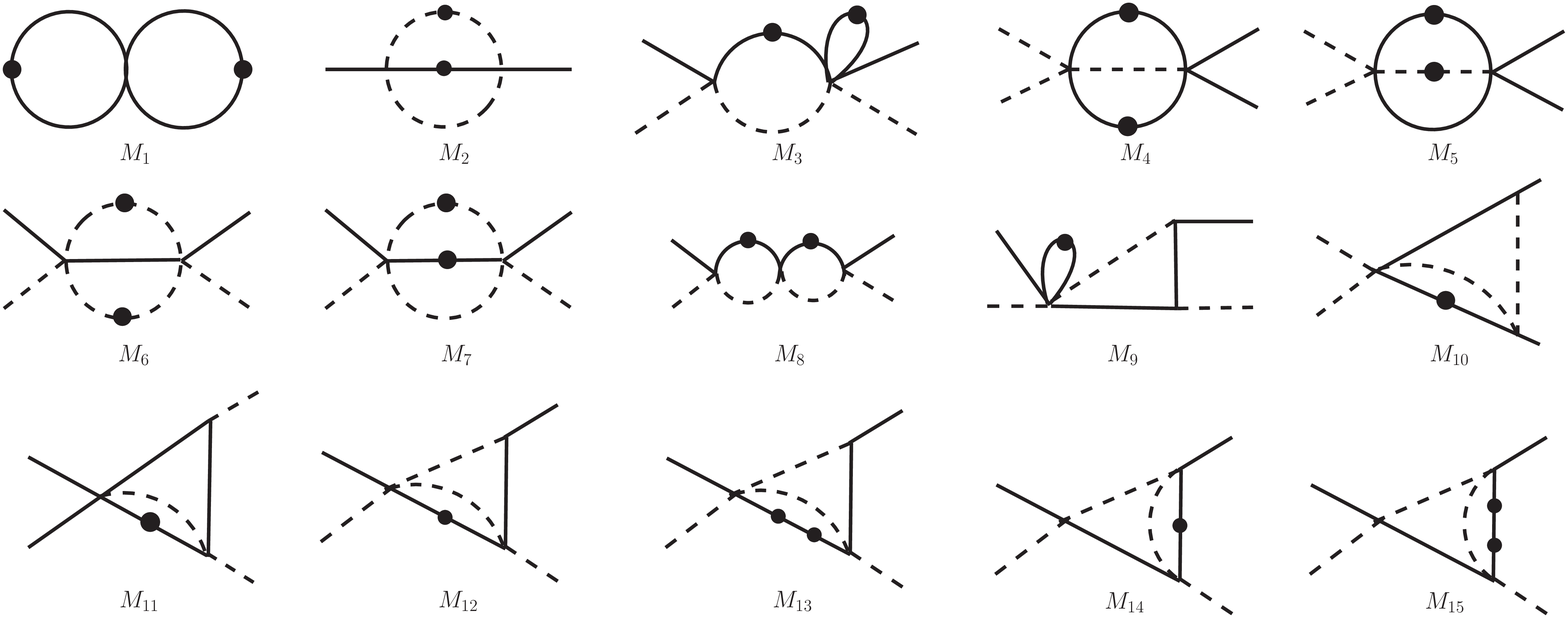}
\includegraphics[scale=0.3]{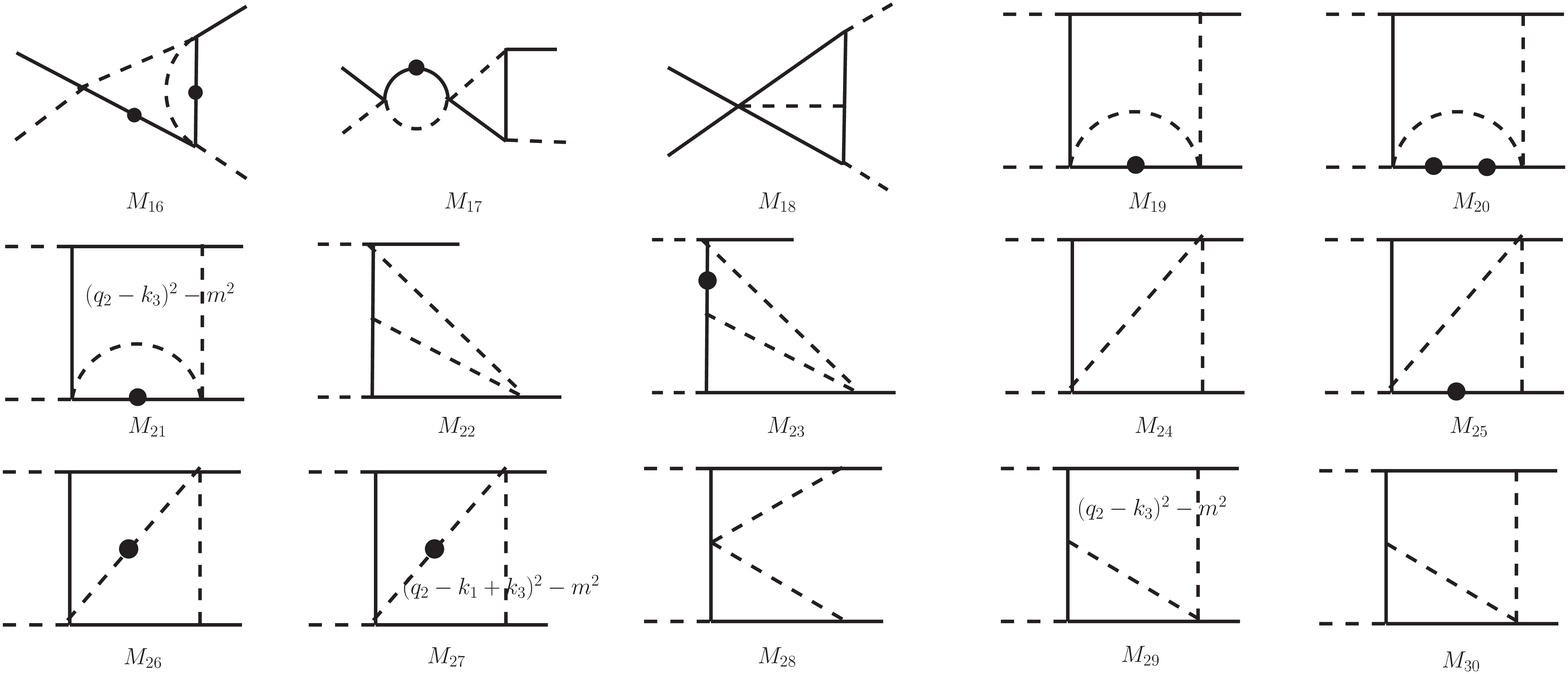}
\includegraphics[scale=0.3]{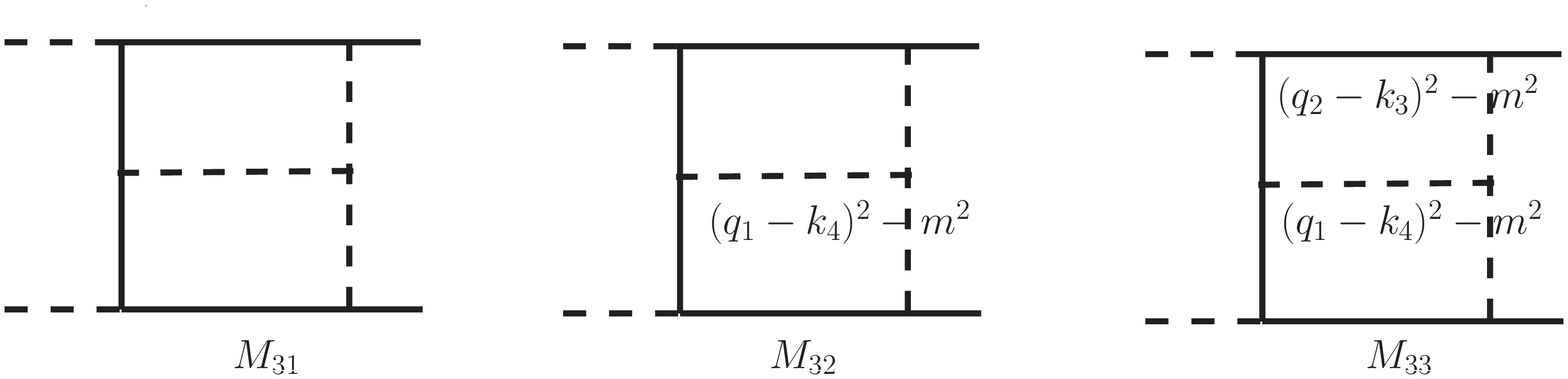}
\caption{  The master integrals for top-quark pair  production shown in Fig.\ref{midiag1}. The solid lines represent massive propagators, while the dashed lines indicate massless ones. 
Each black dot indicates an additional power of the corresponding propagator. For some integrals, we have inserted one or two numerators indicated explicitly on top of the diagram.}
\label{midiag2}
\end{center}
\end{figure}

We have adopted the {\tt FIRE} package \cite{Smirnov:2014hma} to construct the IBP identities. 
All the integrals in this family can be reduced to 66 master integrals. After considering the symmetries of the integrals, e.g.
$(k_1\leftrightarrow k_2, k_3\leftrightarrow k_4, q_1\leftrightarrow q_2)$, there are only 33 master integrals 
that have been shown in Fig.\ref{midiag2}.

As proposed in ref. \cite{Henn:2013pwa}, a proper choice of the basis can lead to a rather simple form of differential equations for the master integrals. In this case the differential equations can be transformed to the $d$-log form, and as a consequence we can present the results of master integrals 
in terms of multiple polylogarithms.
To this aim, we choose the canonical basis as below.
\begingroup
\allowdisplaybreaks[1]
\begin{align*}
\GG_{1}&=   \M_1\,, & \qquad
\GG_{2}&= m^2  \,  \M_2\,,\\
\GG_{3}&= t \,  \M_3\,, & \qquad
\GG_{4}&= \frac{\sqrt{s(s-4m^2)}}{2}\, (2\M_4+\M_5)\,,   \\
\GG_{5}&=- s\, \M_5 \,, &\qquad
\GG_{6}&= (t-m^2)\,\M_6-2m^2 \,  \M_7\,, \\
\GG_{7}&= t\,  \M_7\,,  &\qquad
\GG_{8}&= t^2\,\M_8\,,  \\
\GG_{9}&= (t-m^2)\,  \M_9\,, & \qquad
\GG_{10}&=\sqrt{s(s-4m^2)}\,\M_{10}\,, \\
\GG_{11}&= -s\,\M_{11}\,,& \qquad
\GG_{12}&= (t-m^2)\, \M_{12}\,,   \\
\GG_{13}&= (t-m^2)m^2\, \M_{13}\,,  & \qquad
\GG_{14}&= (t-m^2)  \, \M_{14}\,,\\
\GG_{15}&= (t-m^2)m^2\, \M_{15}\,, & \qquad
\GG_{16}&= (t-2 m^2)m^2 \, \M_{16}-2m^4 \M_{15}-3m^2 \M_{14}\,, \\
\GG_{17}&= t(t-m^2)\, \M_{17}\,, &\qquad
\GG_{18}&= -s \, \M_{18}\,,  \nonumber
\end{align*}
\endgroup
\bqa
\GG_{19}&=& \sqrt{(t-m^2)[t(s-m^2)^2-(s^2-6 s m^2+m^4)m^2]} \, \M_{19}\,,\nonumber\\
\GG_{20}&=& \sqrt{s(s-4m^2)}(t-m^2)\left(\M_{19}+m^2\M_{20}\right)\,,  \nonumber\\
\GG_{21}&=& (t-m^2)\, \left( \M_{21}-m^2 \M_{19}\right)-2(s+t-m^2)\, \M_{11}\,,  \nonumber\\
\GG_{22}&=& (t-m^2) \, \M_{22}\,, \nonumber\\
\GG_{23}&=&  (t-m^2)m^2 \, \M_{23}\,,  \nonumber\\
\GG_{24}&=& (m^2-s-t)\, \M_{24}\,, \nonumber\\
\GG_{25}&=& (t-m^2)m^2\, \M_{25}\,, \nonumber\\
\GG_{26}&=& \sqrt{s(s-4m^2)}(t-m^2) \, \M_{26}\,,  \nonumber\\
\GG_{27}&=& \frac{1}{4}(m^2-s-t)\, (4\M_{27}-2\M_{10}+2\M_{4}+\M_{5}) \,,  \nonumber\\
\GG_{28}&=& (t-m^2)^2 \, \M_{28}\,,  \nonumber
\eqa
\bqa
\GG_{29}&=&(t-m^2)\, (\M_{29}+\M_{24}-\M_{21}+m^2 \M_{19})+2(m^2-s-t)(\M_{18}-\M_{11})\,, \nonumber\\
\GG_{30}&=& \sqrt{s(s-4m^2)}(t-m^2) \, (\M_{30}-m^2 \M_{20}-\M_{19})\,,  \nonumber\\
\GG_{31}&=&   \sqrt{s(s-4m^2)}(t-m^2)^2 \, \M_{31}\, ,\nonumber\\
\GG_{32}&=&  (t-m^2)^2 \, \M_{32}+(t-m^2)(2m^2-s-2t)\, \M_{30}\,,  \nonumber\\
\GG_{33}&=&  t(\M_{33}-4\M_{29}-2\M_{24})-2(t-m^2)m^2\, \M_{32}+2(s+2t-2m^2)m^2\, \M_{30}  \, \nonumber\\
&-&(t(s+t)-2m^2\, t +m^4)\M_{28}+4(s+t)\M_{18}-2m^4\, \M_{25}-2m^4\, \M_{23}\, \nonumber\\
&-&2m^2\, \M_{22}-2m^2\, \M_{21}+2m^4\, \M_{19}+2m^2\, t \,  \M_{17}+4m^4\, \M_{15}+2m^2\M_{14}\, \nonumber\\
&-&2m^4\M_{13}-2m^2\M_{12}+4m^2\M_{11}-2m^2\M_{9}-2m^2\M_{6}-\frac{8m^4}{t-m^2}\M_{2}\, \nonumber\\
&+&\frac{4(t+m^2)m^2}{t-m^2}\, \M_{7} +\frac{t}{2(1-2\epsilon)}(\, \M_{3}+\, \M_{7}-t\, \M_{8})\, 
\eqa
with
\begin{align*}
\M_{1}&=\epsilon^2 \, I_{0, 0, 0, 2, 0, 2, 0, 0, 0}\,,  &
\M_{2}&=\epsilon^2 \, I_{0, 1, 2, 2, 0, 0, 0, 0, 0}\,,  &
\M_{3}&=\epsilon^2 \, I_{1, 0, 0, 0, 2, 2, 0, 0, 0}\,,  \\
\M_{4}&=\epsilon^2 \, I_{0, 0, 2, 2, 0, 1, 0, 0, 0}\,,  &
\M_{5}&=\epsilon^2 \, I_{0, 0, 1, 2, 0, 2, 0, 0, 0}\,,  &
\M_{6}&=\epsilon^2 \, I_{0, 2, 2, 0, 1, 0, 0, 0, 0}\,,  \\
\M_{7}&=\epsilon^2 \, I_{0, 1, 2, 0, 2, 0, 0, 0, 0}\,,  &
\M_{8}&=\epsilon^2 \, I_{1, 1, 0, 0, 2, 0, 2, 0, 0}\,,  &
\M_{9}&=\epsilon^3 \, I_{0, 1, 0, 2, 0, 1, 1, 0, 0}\,,  \\
\M_{10}&=\epsilon^3 \, I_{1, 0, 1, 1, 0, 2, 0, 0, 0}\,,  &
\M_{11}&=\epsilon^3 \, I_{0, 0, 1, 1, 1, 2, 0, 0, 0}\,,  &
\M_{12}&=\epsilon^3 \, I_{0, 1, 1, 0, 2, 1, 0, 0, 0}\,,  \\
\M_{13}&=\epsilon^2 \, I_{0, 1, 1, 0, 3, 1, 0, 0, 0}\,,  &
\M_{14}&=\epsilon^3 \, I_{0, 1, 1, 2, 0, 0, 1, 0, 0}\,,  &
\M_{15}&=\epsilon^2 \, I_{0, 1, 1, 3, 0, 0, 1, 0, 0}\,,  \\
\M_{16}&=\epsilon^2 \, I_{0, 1, 1, 2, 0, 0, 2, 0, 0}\,,  &
\M_{17}&=\epsilon^3 \, I_{1, 1, 0, 0, 2, 1, 1, 0, 0}\,,  &
\M_{18}&=\epsilon^4\, I_{0, 0, 1, 1, 1, 1, 1, 0, 0}\,,  \\
\M_{19}&=\epsilon^3\, I_{0, 1, 1, 2, 0, 1, 1, 0, 0}\,,  &
\M_{20}&=\epsilon^2 \, I_{0, 1, 1, 3, 0, 1, 1, 0, 0}\,, &
\M_{21}&=\epsilon^3 \, I_{0, 1, 1, 2, 0, 1, 1, 0, -1}\, , \\
\M_{22}&=\epsilon^4\, I_{0, 1, 1, 1, 1, 0, 1, 0, 0}\,,  &
\M_{23}&=\epsilon^3 \, I_{0, 1, 1, 1, 1, 0, 2, 0, 0}\,, &
\M_{24}&=\epsilon^4 \, I_{0, 1, 1, 1, 1, 1, 0, 0, 0}\, , \\
\M_{25}&=\epsilon^3\, I_{0, 1, 1, 1, 1, 2, 0, 0, 0}\,,  &
\M_{26}&=\epsilon^3 \, I_{0, 1, 2, 1, 1, 1, 0, 0, 0}\,, &
\M_{27}&=\epsilon^3 \, I_{0, 1, 2, 1, 1, 1, -1, 0, 0}\, , \\
\M_{28}&=\epsilon^4\, I_{1, 1, 0, 1, 1, 1, 1, 0, 0}\,,  &
\M_{29}&=\epsilon^4 \, I_{0, 1, 1, 1, 1, 1, 1, 0, -1}\,, &
\M_{30}&=\epsilon^4 \, I_{0, 1, 1, 1, 1, 1, 1, 0, 0}\, , \\
\M_{31}&=\epsilon^4\, I_{1, 1, 1, 1, 1, 1, 1, 0, 0}\,,  &
\M_{32}&=\epsilon^4 \, I_{1, 1, 1, 1, 1, 1, 1, -1, 0}\,, &
\M_{33}&=\epsilon^4 \, I_{1, 1, 1, 1, 1, 1, 1, -1, -1}\, . & 
\stepcounter{equation}\tag{\theequation}
\label{def:LBasisT1}
\end{align*}

The differential equations of the canonical basis contain two different square roots, i.e.,
\begin{align}
\sqrt{s(s-4m^2)},\quad
\sqrt{(t-m^2)[t(s-m^2)^2-(s^2-6 s m^2+m^4)m^2]}.
\end{align}
Notice that only $\GG_{19}$ contains the second square root explicitly.
In order to solve the differential equations in terms of multiple polylogarithms,
we have to rationalize these square roots.
Therefore, we define two dimensionless variables,
\beq
y=-\frac{\sqrt{s}-\sqrt{s-4m^2}}{\sqrt{s}+\sqrt{s-4m^2}}\,,\qquad
x=(y^2-y+1)\frac{\sqrt{t-m^2(s^2-6s\, m^2+m^4)/(s-m^2)^2}}{\sqrt{t-m^2}}\,
\eeq
corresponding to the following transformation of variables
\beq
s=-\frac{(1-y)^2}{y}\, m^2\,, \qquad t=\frac{x^2-(y-1) y [y (y+3)-2]-1}{x^2-((y-1) y+1)^2}\, m^2 \,.
\eeq
After changing to the $x$ and $y$ variables, the differential equations for $\text{{\bf F}}=\{ \GG_1,\ldots ,\GG_{33}\}$  can be written as
\bqa
d\, \text{{\bf F}}(x,y;\epsilon)=\epsilon\, (d \, {\bf A})\,  \text{{\bf F}}(x,y;\epsilon),
\eqa
where
\bqa
d\,{\bf A}=\sum_{i=1}^{13} \text{\bf R}_i\,  d \log(l_i)
\label{eq:dA}
\eqa
with $\text{\bf R}_i$ rational matrices independent of the kinematics and the space-time dimension.
Their explicit forms are provided in ancillary files.
These $d$-log forms contain all the information of the kinematics.
The set of the arguments $l_i$ is referred to as the {\it alphabet} and it consists of the following 13 letters
\begin{align}
\begin{alignedat}{2}
l_1 & =x-(y^2+y-1)\,,&\quad
l_2 & =x+(y^2+y-1)\,, \\
l_3 & =x-(y^2-y-1)\,, &\quad
l_4 & =x+(y^2-y-1)\,, \\
l_5 & =x-(y^2-y+1)\,,&\quad
l_6 & =x+(y^2-y+1)\,,\\
l_7 & =x-(y^2-3 y+1)  \,,&\quad
l_8 & =x+(y^2-3 y+1) \, , \\
l_9 & =x^2-[y (y-3) +1] \left(y^2+y+1\right)\,,&\quad
l_{10} & =x^2-y (y-1) [y (y+3)-2]-1\, , \\
l_{11} & =y\,,&\quad
l_{12} & =y+1\, , \\
l_{13} & =y-1\,.
\end{alignedat} \stepcounter{equation}\tag{\theequation}
\label{alphabet}
\end{align}

One can see from Fig.\ref{midiag2} that the master integrals
 $\{\M_{29},\M_{30},\M_{31},\M_{33}\}$ contain $\M_{19}$ as a sub-topology.
As a consequence, their differential equations may contain $\M_{19}$.
However, we choose the canonical basis  $\{\GG_{29},\GG_{30},\GG_{31},\GG_{33}\}$
in such a way that their dependence on  $\GG_{19}$ has been removed.
At the end, only the differential equations of $\{\GG_{19},\GG_{20},\GG_{21},\GG_{32}\}$ involve $\GG_{19}$. 

The differential equations of the remaining integrals do not depend on $\GG_{19}$ and thus do not contain the square root $\sqrt{(t-m^2)[t(s-m^2)^2-(s^2-6 s\, m^2+m^4)m^2]}$. Since the result of $\GG_{19}$ starts at  transcendental  weight three as indicated by Eq.(\ref{eq:dA}), we can use the  variables $y$ and $z\equiv t/m^2$, instead of $y$ and $x$, to express the results of the remaining integrals,
namely all the master integrals except $\{\GG_{19},\GG_{20},\GG_{21},\GG_{32}\}$, up to  transcendental weight four.
In this way, we obtain more compact results for the remaining master integrals.
Here, for illustration, we show the differential equations for $\GG_{31}$, 
\bqa
\frac{\partial\, \GG_{31}}{\partial\, y}&=&\epsilon\bigg[\frac{1}{y-z}(4 \text{F}_2+\text{F}_6-2 \text{F}_7+\text{F}_9+3 \text{F}_{12}+4 \text{F}_{13}-4\text{F}_{14}-6 \text{F}_{15}-2 \text{F}_{17}+2 \text{F}_{20}\nonumber\\&+&2 \text{F}_{22}+2\text{F}_{23}-\text{F}_{26}+2 \text{F}_{31}-2 \text{F}_{32})-\frac{1}{y-\frac{1}{z}}(4 \GG_2+\GG_6-2 \GG_7+\GG_9+3 \GG_{12}\nonumber\\&+&4 \GG_{13}-4 \GG_{14}-6 \GG_{15}-2 \GG_{17}-2 \GG_{20}+2 \GG_{22}+2 \GG_{23}+\GG_{26}-2\GG_{31}-2\GG_{32})\nonumber\\&+&\frac{1}{2 y}(5 \text{F}_1+36 \text{F}_2-4 \text{F}_3-2 \text{F}_5+4 \text{F}_6-12\text{F}_7+8\text{F}_9-10\text{F}_{11}+8\text{F}_{12}+8\text{F}_{13}\nonumber\\
&-&24\text{F}_{14}-24\text{F}_{15}-8\text{F}_{16}-8 \text{F}_{17}-20 \text{F}_{18}+8 \text{F}_{20}+8\text{F}_{22}+8 \text{F}_{23}+16 \text{F}_{24}-4 \text{F}_{26}\nonumber\\
&+&4 \text{F}_{27}-4\text{F}_{28}-8 \text{F}_{29}+4 \text{F}_{31}+8 \text{F}_{32}-8 \text{F}_{33})-\frac{2}{y-1}(2 \text{F}_{20}-\text{F}_{26}+\text{F}_{31})\nonumber\\&-&\frac{2}{y+1}(4 \text{F}_{20}-2 \text{F}_{26}+3 \text{F}_{31})\bigg]\,,\nonumber
\eqa
\bqa
\frac{\partial\, \GG_{31}}{\partial\, z}&=&\epsilon\bigg[\frac{1}{z-y}(4 \GG_2+\GG_6-2 \GG_7+\GG_9+3 \GG_{12}+4 \GG_{13}-4 \GG_{14}-6 \GG_{15}-2 \GG_{17}+2 \GG_{20}\nonumber\\
&+&2 \GG_{22}+2\GG_{23}-\GG_{26}+2 \GG_{31}-2 \GG_{32})-\frac{1}{z-\frac{1}{y}}(4 \GG_2+\GG_6-2 \GG_7+\GG_9+3 \GG_{12}\nonumber\\&+&4 \GG_{13}-4 \GG_{14}-6 \GG_{15}-2 \GG_{17}-2 \GG_{20}+2 \GG_{22}+2 \GG_{23}+\GG_{26}-2 \GG_{31}-2\GG_{32})\nonumber\\&-&\frac{4\GG_{31}}{z-1}\bigg]\,.
\label{eq:F31}
\eqa

\section{Boundary conditions and analytic results}
\label{sec3}

In order to obtain analytic results from differential equations for the canonical basis shown in the previous section, we need to determinate the boundary conditions first.

The bases $\{ \GG_1,\GG_2\}$ are just single-scale integrals, corresponding to a vacuum diagram with virtual massive particles or a self-energy diagram of a massive particle, and their results are already known in ref. \cite{Chen:2017xqd}. 
Explicitly, 
\beq
\GG_1=1\, , \, \GG_2=-\frac{1}{4}-\epsilon^2\frac{\pi^2}{6}-2\epsilon^3\zeta(3)-\epsilon^4\frac{8\pi^4}{45}+{\cal O}(\epsilon^{5}).
\label{eq:F2}
\eeq
The boundary condition for $\GG_6$ at $t=0~(z=0)$ can be evaluated 
using the Mellin-Barnes method, implemented in the Mathematica packages {\tt MB} \cite{Czakon:2005rk} and {\tt AMBRE} \cite{Gluza:2007rt}, 
and we obtain
\beq
\GG_{6}|_{ z=0 }=1+\epsilon^2\frac{\pi^2}{3}-2\epsilon^3\zeta(3)+\epsilon^4\frac{\pi^4}{10}+{\cal O}(\epsilon^{5}).
\label{eq:F6}
\eeq

All the master integrals do not have singularities at $t=0$. Due to the prefactor $t$ for the bases $\{\GG_3,\GG_7,\GG_8,\GG_{17}\}$, we can deduce that they are vanishing at $t=0$. 
We derive the boundary of $\GG_{16}$ at $t=0$ from the regularity condition of the corresponding differential equation at $t=0$. Specifically, the differential equation for $\GG_{16}$ can be formulated as 
\beq
\frac{d \GG_{16}}{d t}=-\frac{\epsilon}{2t}(3\GG_1+12\GG_2-2\GG_3-6\GG_{14}-6\GG_{16})+\ldots,
\eeq
where the ellipses stand for less singular terms at $t=0$. The regularity condition at $t=0$ leads to a relation 
\beq
\lim_{t\rightarrow 0} (3\GG_1+12\GG_2-2\GG_3-6\GG_{14}-6\GG_{16})=0.
\eeq
Thus, we obtain the boundary condition of $\GG_{16}$ at $t = 0~(z=0)$ from the above equation.

All the master integrals are regular at $s=0$.
The canonical bases $\{\GG_4,\GG_5,\GG_{10},\GG_{11},\GG_{18},\GG_{31}\}$ have a prefactor $s$ or $\sqrt{s(s-4m^2)}$
 and thus they  are vanishing at $s=0$. The boundary condition of $\GG_{33}$ is determined from the regularity condition of the corresponding differential equation at $s=0~(y=1)$. 
 
The same logic leads us to know that the bases $\{\GG_9,\GG_{12},\GG_{13},\GG_{14},\GG_{15},\GG_{22},\GG_{23},\GG_{25},\GG_{28}, \GG_{30}\}$ are vanishing at $t=m^2$,
and $\GG_{29}=2\GG_{18}-2\GG_{11}$ at $t=m^2~(z=1)$.

The base $\GG_{19}$ does not have a singularity at $t=\frac{(s^2-6 s m^2+m^4)m^2}{(s-m^2)^2}$ but has a prefactor $\sqrt{t(s-m^2)^2-(s^2-6 s m^2+m^4)m^2}$, so it is vanishing at $t=\frac{(s^2-6 s m^2+m^4)m^2}{(s-m^2)^2}$ $(x=0)$. 

The planar integrals we are considering do not have a $u$-channel singularity.  Due to the prefactor $(m^2-s-t)=(u-m^2)$ for $\{\GG_{24},\GG_{27}\}$, they are vanishing at $t=m^2-s$.

The boundary conditions of $\{\GG_{20},\GG_{21},\GG_{26},\GG_{32},\GG_{33}\}$ are determined from the regularity conditions of the corresponding differential equations at $t=\frac{m^4}{u}(z=y)$.

With the discussion above, we finish the determination of all boundary conditions that are necessary to obtain the full analytic results.
Then all the master integrals can readily be calculated from the differential equations. The analytic results for $\{\GG_1,\ldots, \GG_{33}\}$ up to transcendental  weight four are expressed in terms of multiple polylogarithms \cite{Goncharov:1998kja}, which  are provided in an ancillary file. For illustration, we show the result for $\GG_{33}$ up to transcendental  weight four
\bqa
\GG_{31}&=&\epsilon^3\bigg[2G_{0,0,0}(y)+\frac{\pi^2}{3}G_{0}(y)\bigg]+\epsilon^4\bigg[-\frac{5\pi^4}{18}+\frac{\pi^2}{3}(2G_{1,0}(y)+2G_{-1,0}(y)-7G_{0,0}(y))\nonumber\\
&+&4G_{-1,0,0,0}(y)-12G_{0,0,-1,0}(y)-4G_{0,0,0,0}(y)+4G_{0,0,1,0}(y)+8G_{0,1,0,0}(y)\nonumber\\
&+&4G_{1,0,0,0}(y)-8G_{0,0,0}(y)G_1(z)+\frac{2}{3}G_{0}(y)[\pi^2(3G_{0}(z_1)-2G_{\frac{1}{z}}(z_1)-2G_{1}(z))\nonumber\\
&-&6(G_{\frac{1}{z},0,0}(z_1)-G_{\frac{1}{z},1,0}(z_1)+G_{z,1,0}(z_1)+G_{1,0,1}(z)+G_{\frac{1}{z},1,0}(1)-G_{z,1,0}(1)\nonumber\\
&+&2G_{0,1,0}(z_1)+3G_{1,0,0}(z_1)-G_{0,0,1}(z)-3G_{0,0,0}(z_1))-9\zeta(3)]\bigg]+{\cal O}(\epsilon^{5}),
\eqa
where $z_1=(z+1+\sqrt{z^2+2z-3})/2$ corresponds to the boundary condtions of $F_{24},F_{27}$ discussed above. 
Here $G_{a_1,a_2,...,a_n}(x)$ are multiple  polylogarithms \cite{Goncharov:1998kja}, defined recursively by
\begin{align}
    G_{a_1,a_2,...,a_n}(x) = \int _0 ^x \frac{dt}{t-a_1}G_{a_2,...,a_n}(t)
\end{align}
with $G(x)=1$ and
\begin{align}
    G_{\underbrace{0,0,...,0}_{n}}(x) = \frac{1}{n!}\ln ^n x\,.
\end{align}
The number $n$ is referred to as the  transcendental weight of the result.
We see that the result of $F_{31}$ begins with $O(\epsilon^3)$. This is anticipated.
From the discussion above, we see that the only nonvanishing boundary values are that of $F_{6}$ in Eq.(\ref{eq:F6}) and the two single-scale integrals $F_{1}, F_{2}$ in Eq.(\ref{eq:F2}).
Though they have $O(\epsilon^0)$ contribution, 
their combination in Eq.(\ref{eq:F31}) is in such a way that the right-hand side starts at $O(\epsilon^3)$.

The multiple polylogarithms can be numerically evaluated by the {\tt GINAC} implementation \cite{Vollinga:2004sn,Bauer:2000cp}. They can also be transformed to the functions like   $ \text{Li}_n(x)$ and $\text{Li}_{2,2}(x,y)$  up to  transcendental weight four with the methods described in \cite{Frellesvig:2016ske}.

We are interested in the physical region for top-quark pair production, i.e., $(s>4m^2,t<-s(1-\sqrt{1-4m^2/s})^2/4)$.  The proper analytic continuation can be achieved by starting from the Euclidean space where $s<4m^2$ and $t<0$.
Then we assign $s$ a small positive imaginary part $(s\rightarrow s+0 i)$ to produce the correct result when $s>4m^2$ \cite{Gehrmann:2002zr}.

The analytic results of all the master integrals have been checked with the numerical package {\tt FIESTA} \cite{Smirnov:2015mct},
and good agreements have been achieved.
For example, we show the result of $\text{M}_{31}$ at a  kinematic point   $(s=3.41,t=-0.91,m=1)$,
\bqa
\text{M}_{31}^{\rm analytic}&=&\epsilon^3(0.656683)\,\,\,\,\,\,\,\,\,\,\,\,\,\,\,+\,\,\,\,\,\,\,\,\,\,\,\,\,\,\,\epsilon^4(4.006860),\nonumber\\
\text{M}_{31}^{\rm FIESTA}&=&\epsilon^3(0.656684\pm0.000002)+\epsilon^4(4.006867 \pm 0.000019).
\eqa

\section{Conclusions}
\label{sec4}

In summary, we calculate analytically a planar double-box Feynman integral family for top-quark pair production.
After choosing a canonical basis, the differential equations for the corresponding basis are expressed in canonical form.
The boundary conditions are determined either by simple integrals
or by regularity conditions at certain kinematic points without physical singularities.
Therefore, the analytic results of the basis can be expressed in terms of multiple polylogarithms. 
These results and the rational matrices in the differential equations are provided in ancillary files. 
In the future, it will be interesting to calculate
the other unknown integral families,
especially those with elliptic integrals,
to obtain a fully analytic result of the two-loop
corrections to the top-quark pair production.

\section*{Acknowledgments}
\noindent This work was supported by the National Natural Science Foundation of China(NSFC) under the grants 11747051 and 11805042.
The work of J.W was supported by the BMBF project No. 05H15WOCAA and 05H18WOCA1. 

\bibliography{dtop}
\bibliographystyle{JHEP}

\end{document}